# LOCALIS: Locally-adaptive Line Simplification for GPU-based Geographic Vector Data Visualization

Alireza Amiraghdam†, Alexandra Diehl, Renato Pajarola

Department of Informatics, University of Zürich, Switzerland

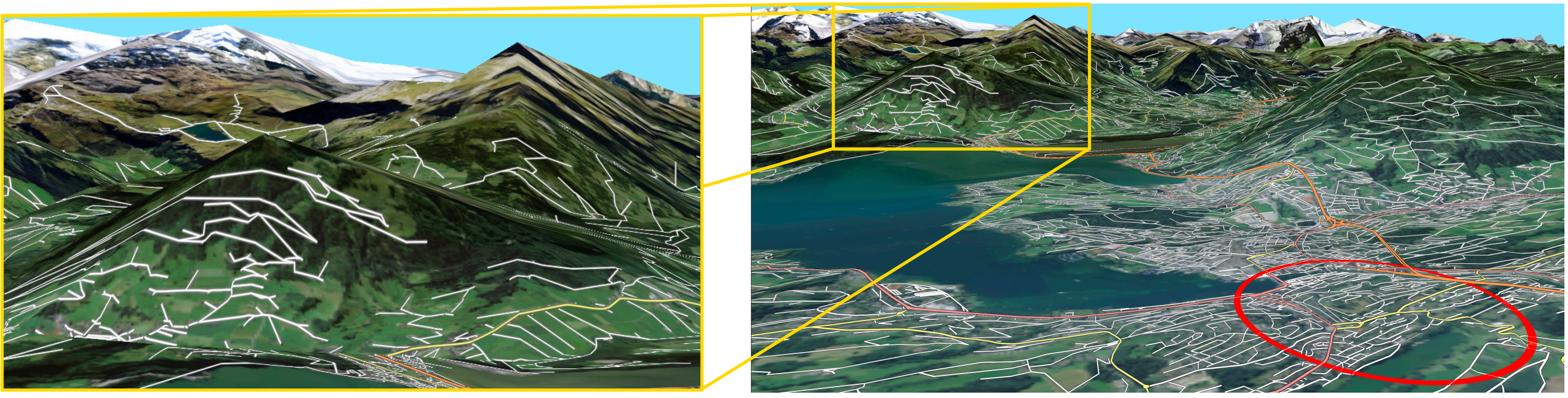

**Figure 1:** *Street network visualized over a textured terrain model using* LOCALIS. *Left zoom-in inset highlights the distance based simplification of lines. Red circle on the right demonstrates the flexibility of the approach using a line-refinement lens that can simplify or refine lines inside a region interactively controlled by the user.*

**Abstract**

*Visualization of large vector line data is a core task in geographic and cartographic systems. Vector maps are often displayed at different cartographic generalization levels, traditionally by using several discrete levels-of-detail (LODs). This limits the generalization levels to a fixed and predefined set of LODs, and generally does not support smooth LOD transitions. However, fast GPUs and novel line rendering techniques can be exploited to integrate dynamic vector map LOD management into GPU-based algorithms for locally-adaptive line simplification and real-time rendering. We propose a new technique that interactively visualizes large line vector datasets at variable LODs. It is based on the Douglas-Peucker line simplification principle, generating an exhaustive set of line segments whose specific subsets represent the lines at any variable LOD. At run time, an appropriate and view-dependent error metric supports screen-space adaptive LOD levels and the display of the correct subset of line segments accordingly. Our implementation shows that we can simplify and display large line datasets interactively. We can successfully apply line style patterns, dynamic LOD selection lenses, and anti-aliasing techniques to our line rendering.*

## 1. Introduction

Interactive visualization of large geographic vector map data is a challenging problem, in particular in combination with real-time adaptive level-of-detail (LOD) methods. LOD-based simplification and rendering techniques offer well-proven solutions for dynamically adjusting the amount and resolution of the data to be displayed. In the context of scientific visualization and computer graphics, the development of multiresolution LOD methods has been an active research area that resulted in many algorithms and data structures to facilitate real-time rendering of very large amounts of 3D data, e.g. such as polygonal meshes, volumes or point cloud data.

For 2D textures, 3D meshes or volumetric data, different LOD

† Authors emails: {amiraghdam,diehl,pajarola}@ifi.uzh.ch

simplification algorithms for both off-line preprocessing and real-time visualization have been proposed. In contrast, online LOD management and LOD-based interactive visualization of vector map line data has not received the same amount of attention. In 3D rendering engines, polyline data is commonly dealt with by being transformed into other formats such as textures or meshes, and the corresponding LOD techniques for these formats are then applied. In *geographic information systems* (GIS), besides terrain elevation models and texture maps, large parts of the particularly important cartographic data is given in vector format.

Cartographic generalization techniques for vector data have been studied for decades. However, vector map data processing and interactive visualization have typically been handled independently, thus no efficient and fully integrated solutions have been proposed until now. Furthermore, the performance of prior online methods is still insufficient for real-time line simplification and display of larger datasets, and therefore, past approaches do not easily translate to dynamic interactive visualizations. In particular, the outcome of the polyline generalization has not been tailored towards interactive 3D visualization. As a consequence, there is a lack of real-time adaptive LOD techniques for vector polyline data.

We tackle the lack of specific techniques for visualizing vector polylines. Moreover, we consider line features to be displayed not only in 2D but in an interactive 3D geographic visualization. Following recent achievements for fast vector map rendering in 3D [TBP16, TBP18, FEP18], we propose a novel algorithm, *locally-adaptive line simplification* (LOCALIS), for GPU-based geographic polyline data visualization. Our main contributions are (1) a GPU-based view-dependent Douglas-Peucker style polyline simplification approach, that exploits (2) a novel LOD line-segments data structure, and (3) an efficient GPU-based deferred line rendering algorithm. Furthermore, in the experimental results, we not only demonstrate the interactive LOD simplification and rendering performance of our approach but also its line-styling as well as screen-space LOD and data filtering features. We focus on simple static line data, i.e. polylines that only intersect at joints and which do not have further temporal attributes such as e.g. trajectories. Our technique may work with other lines or trajectories but does not take their specific error metrics into account.

## 2. Related Work

### 2.1. Geographical Vector Data Visualization

In 3D, vector line data is most commonly displayed as an overlay over a terrain model, which can be a simple flat plane with a texture resembling the terrain as a 2D image, or it can be a digital 3D elevation model. In this context, we can categorize the methods for rendering geographic vector data into four major groups: (1) texture-based overlays, (2) geometry-based methods, (3) shadow-volume-based techniques, and (4) deferred direct vector rendering.

In texture-based methods [KD02, WKW*03, SLL08, WLB09], vector data is rasterized and stored as a texture which is then projected over the terrain during rendering. These methods are fast and easy to implement but suffer from an insufficient resolution in areas closer to the camera, aliasing artifacts in far areas as well as projective distortions. To overcome these problems, higher or multi-resolution textures are used at the expense of larger texture memory usage. Furthermore, view-dependent and dynamically adapting vector maps require the textures to be updated each frame.

Geometry-based approaches, transform the vector data into auxiliary meshes modifying them to match the underlying terrain [QWS*11, WSFL10]. While not suffering from resolution problems, other drawbacks arise. First, creating meshes from large-scale vector maps results in an even larger amount of geometry to be rendered as each line primitive gives rise to several polygon primitives. Another issue is matching the meshes to the 3D terrain, especially in connection with multiresolution view-dependent LOD visualization approaches in which the terrain mesh as well as the vector maps continuously change as the camera moves. Furthermore, like texture based solutions, these auxiliary meshes have to be recreated whenever the vector maps change, in the worst case before each frame. In general, unpredictable scene configurations are problematic for geometry-based line rendering methods.

In shadow-volume techniques, the vector map polylines are considered floating above the terrain and orthogonal shadow polygons are created intersecting the terrain [DZY08, YZK*10]. The advantage of this method is the independence of vector data from the terrain model. However, it does not scale well and requires multiple geometry rendering passes. For large vector datasets, a high number of shadow volumes must be created resulting in expensive shadow computations. In case of dynamically changing vector maps, shadow volumes must be updated every frame.

We based our approach on the deferred vector map rendering [TBP16, TBP18, FEP18]. In this technique, the vector line data is maintained on the GPU in a data structure that allows fast screen-space (pixel) to object-space (line) search and mapping. Moreover, the search for lines close to a given pixel can be done efficiently by using this GPU-based vector map data structure, and the pixel can be colored based on the distance from the line. The advantage of this approach is that there is no loss of precision or distortion of the vector map since there is no intermediate transformation into textures, meshes or shadow volumes. Therefore, the accuracy of the final result is as high as the resolution of the output allows. A more comprehensive list of the drawbacks and advantages of the different group of methods can also be found in [TBP18].

### 2.2. Cartographic Generalization

Generalization is a key concept in cartography and has been used for displaying maps at different scales, with the goal of adjusting the amount and visual complexity of cartographic elements to match a specific use case and spatial resolution. Such generalized maps are supposed to simplify a given task and increase the efficiency of the users [WBW10]. Generalization is done by applying different operations to cartographic elements which are classified into several categories such as elimination, simplification, aggregation, and collapse [MS92, FSK07, RBS11].

Automated line simplification and feature selection methods help to reduce time-consuming manual work and maintain consistency [BW88]. Early batch processing [HW07] methods worked by chaining several operations sequentially and providing the necessary control parameters. Subsequent improvements included rule-based expert systems which modeled cartographic generalization knowledge as a set of rules [BM91]. Due to the complexity of the generalization process, a high number of rules were needed. In addition, as the number of rules increased, new problems emerged such as conflicts and competitions between rules [FM87]. Eventually, expert systems can be used for specific problems such as label placement, but are getting too complex for the whole process of generalization [Zor91].

The constraints concept [Bea91] defines the desired output by constraints, and an algorithm optimizes the combination of generalization operations in order to produce the best output based on the defined constraints. Among the optimization techniques that have been developed for this purpose, the agent-based method [LRD*99] has successfully been used in map production [RRB11]. However, this approach is still not effective enough for on-demand map generalization because defining constraints for every possible situation that users could demand is not possible. To try to overcome this shortcoming, ontology-based approaches were proposed for road line simplification [KDE05] and road accident visualization [GM16]. Still, a comprehensive ontology has not been created to cover the whole generalization process.

Early interactive map visualization systems used a set of maps at different discrete cartographic scales which can be selected based on the user interaction and display resolution. To avoid the limitations of using only a given set of discrete LODs, on-the-fly generalization approaches keep the vector map in data structures that can be used to extract maps at a desired detail level on demand [WB08]. With respect to linear vector map features, the *binary line generalization* (BLG) tree [VOVDB89] is an important basic line simplification data structure based on Douglas-Peucker (DP) algorithm. Reactive-trees [VO92] as well as generalized area partitioning trees [VO95] were designed for on-the-fly line simplification, as well as the Multi-VMap [VMPR06].

Despite these advances, cartographic vector map line generalizations are still far away from real-time performance on larger scales and are not considering interactive 3D visualization scenarios. Our method, while being limited to line features, is to the best of our knowledge the first real-time locally-adaptive line simplification and visualization solution.

### 2.3. LOD Simplification and Error Metrics

Simplification and multiresolution modeling techniques are widely known for various 3D geometry data types [LRC*03]. Specifically, the general concept of multiresolution hierarchy for view-dependent LOD rendering with a dynamically adapting rendering front (see e.g. also [PD04, HSH10, DVS03]) is also followed in this work, but nevertheless realized in a very different and novel way.

While a variety of error metrics are available for line simplification [ZDY*18, vKLW18], due to its excellent accuracy [SC06] and simplicity, we based our approach on the DP technique and the error metric specific to this technique. For efficient view-dependent and screen-space adaptive LOD selection, we adopt the concept of *error saturation* known from terrain rendering [LP01, BPS04, PG07]. This allows us to define a BLG-tree supporting view-dependent line-refinement operations as described in Section 3.2.

## 3. Locally Adaptive Line Simplification

### 3.1. Douglas-Peucker Line Refinement Trees

Our line simplification approach is based on the DP technique and the BLG-tree [VOVDB89]. Fig. 2 illustrates the DP line refinement principle and the corresponding BLG-tree. The process is defined by incrementally refining the current line version, initially starting with a straight connection between the endpoints. In each step, one line segment acts as a *baseline* which is subdivided by adding a refinement point, and this next point is chosen as the one having the largest distance from its baseline.

The distance $e$ of a point $\boldsymbol{p}_i$ to its baseline is considered to be the *error* that this point introduces when leaving it out for representing the line. Adding a $\boldsymbol{p}_i$ thus causes dividing the corresponding baseline $\overline{\boldsymbol{p}_l\boldsymbol{p}_r}$ and hence also splitting the remaining unused points belonging to the same baseline into two groups $\mathcal{L}_i = \{\boldsymbol{p}_j | l < j < i\}$ and $\mathcal{R}_i = \{\boldsymbol{p}_j | i < j < r\}$. The two sets $\mathcal{L}, \mathcal{R}$ define the two subtrees in the BLG-tree of node $\boldsymbol{p}_i$. In Fig. 2, starting with the initial baseline $\overline{\boldsymbol{p}_B\boldsymbol{p}_E}$, inserting $\boldsymbol{p}_2$ splits the remaining unused points into the sets $\mathcal{L} = \{\boldsymbol{p}_1\}$ and $\mathcal{R} = \{\boldsymbol{p}_3, \ldots \boldsymbol{p}_6\}$. Subsequently, the two new baselines $\overline{\boldsymbol{p}_B\boldsymbol{p}_2}$ and $\overline{\boldsymbol{p}_2\boldsymbol{p}_E}$ with their respective unused points $\mathcal{L}$ and $\mathcal{R}$ are processed. At any moment, there is thus a set of baselines, each with its refinement points, and for the next refinement step the baseline with the refinement point with the largest distance is subdivided, until the desired or the full LOD is reached.

As can be seen in Fig. 2, a binary BLG-tree is built according to the above outlined process. We use this tree to simplify a line adaptively based on a given error threshold. In Fig. 2 for example, if a recursive tree traversal stops when the error $e$ of a node becomes less than 10, the so far traversed tree and selected nodes with $e > 10$ would result in the line $\overline{\boldsymbol{p}_B\boldsymbol{p}_2\boldsymbol{p}_3\boldsymbol{p}_E}$. If instead we use $e > 5$ then the resulting refined line is $\overline{\boldsymbol{p}_B\boldsymbol{p}_1\boldsymbol{p}_2\boldsymbol{p}_3\boldsymbol{p}_5\boldsymbol{p}_E}$.

Fig. 3(a) illustrates the situation in which the error threshold is adaptively defined based on the distance from a camera. A function $\varepsilon(d)$ translates the distance $d$ from the camera to an error threshold that is used to compare the error $e_i$ of each point $p_i$ in the BLG-tree. While traversing the BLG line-refinement tree we thus check for the inequality

$$e_i > \varepsilon(d_i), \quad (1)$$

given the function $\varepsilon()$ which corresponds to a screen-space error threshold. In this equation, $e_i$ denotes the error of the node and $d_i$ its distance to the camera. Each $p_i$ for which this inequality is true is included and refines its baseline, and the recursive traversal stops when the test fails. If the test fails, e.g. at node $\boldsymbol{p}_5$ in Fig. 3(b), the entire subtree is not included for line refinement, and the final result is $\overline{\boldsymbol{p}_B\boldsymbol{p}_1\boldsymbol{p}_2\boldsymbol{p}_3\boldsymbol{p}_E}$ as shown in Fig. 3(c).

### 3.2. GPU-based Line Simplification

The deferred line rendering algorithm described below in Section 3.6 requires efficient *pixel-on-line evaluations*. To avoid many expensive BLG-tree traversals for all pixels, we thus propose a novel approach that converts the line refinement trees into an exhaustive set of *attributed line segments*, and indexes them.

These attributed line segments include all possible line configurations, e.g. as illustrated in Fig. 2, that could be needed for any LOD refinement situation. Furthermore, they are constructed such that their LOD visibility can be determined individually according to a given error threshold. To achieve this, for each BLG-tree we generate the set $\mathcal{T}$ of all possible line segments that can occur by traversing the tree. $\mathcal{T}$ can be extracted from a BLG-tree using the following three rules which are illustrated in Fig. 4, along with an example. Rule 1: connecting $\boldsymbol{p}_B$ and $\boldsymbol{p}_E$ to the root node and treat the root as the right child of $\boldsymbol{p}_B$ and left child of $\boldsymbol{p}_E$. For the next rules, we denote the descendants of a node by their relative paths in the subscript such that the left child of the node $\boldsymbol{p}_i$ is $\boldsymbol{p}_{il}$ and the right child of the right child of the left child of $\boldsymbol{p}_i$ is $\boldsymbol{p}_{ilr^2}$. Rule 2: connecting each descendant $\boldsymbol{p}_{ilr^t}$ with $t > 0$ to $\boldsymbol{p}_i$. Rule 3: Symmetry of Rule 2 by swapping $l$ and $r$.

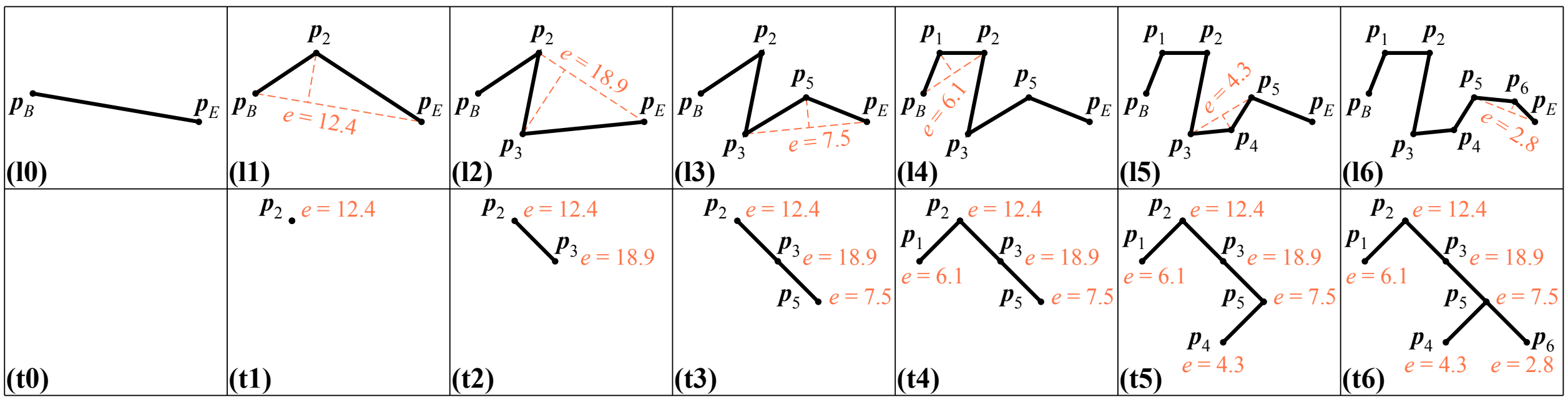

**Figure 2:** *Different steps of the DP algorithm when refining a line (top) and the corresponding tree structure (bottom). (l0) Shows the most simplified and (l6) the fully refined line version. (l1-l5) Indicate the steps where the most impactful point is added each time, with the distances to the subdivided baseline also illustrated. (t0-t6) Show the nodes corresponding to the inserted refinement points being added to an incrementally growing binary tree.*

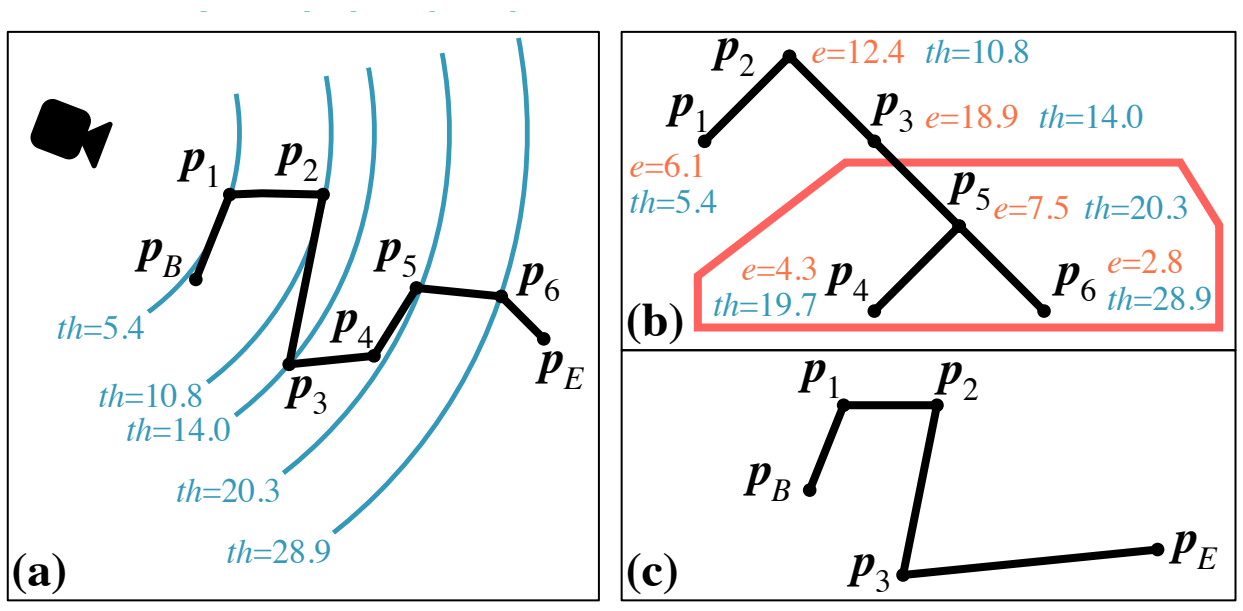

**Figure 3:** *Example with line refinement error dependent on the distance to the camera. (a) For each point $p_i$ an error threshold th is calculated using the given function $\varepsilon(d_i)$. (b) While performing an in-order tree traversal, the error of each node is tested against the calculated threshold. At $p_5$, $e < th$, thus $p_5$ and its children are discarded. (c) The final refined line, with parts closer to the camera having more detail.*

In order to determine the LOD visibility of each individual line segment in $\mathcal{T}$, we need to check the inclusion of its two endpoints. Let us consider the line segment $\overline{\boldsymbol{p}_B\boldsymbol{p}_2}$ in Fig. 2 which appears for the first time in Fig. 2(l1) where $\boldsymbol{p}_2$ is used to refine the line. We can observe that $\overline{\boldsymbol{p}_B\boldsymbol{p}_2}$ will continue to be a part of the simplified line until $\boldsymbol{p}_1$ is included in Fig. 2(l4). The visibility of the line segment $\overline{\boldsymbol{p}_B\boldsymbol{p}_2}$ is thus not affected by any other point. Therefore, we call $\boldsymbol{p}_2$ the *generator* and $\boldsymbol{p}_1$ the *splitter*. Since the generator $\boldsymbol{p}_g$ is always one of the two endpoints, the other being $\boldsymbol{p}_l$, we additionally only need to know the splitter $\boldsymbol{p}_s$ of a line segment. Thus three points $\boldsymbol{p}_l, \boldsymbol{p}_g, \boldsymbol{p}_s$ and their errors $e_l, e_g, e_s$ need to be known and compared to the LOD error threshold to determine the visibility of a specific line segment in $\mathcal{T}$. Therefore, *a line segment is visible when its generator is included and its splitter is not*, i.e. iff

$$e_g > \varepsilon(d_g) \wedge e_s \ngtr \varepsilon(d_s). \qquad (2)$$

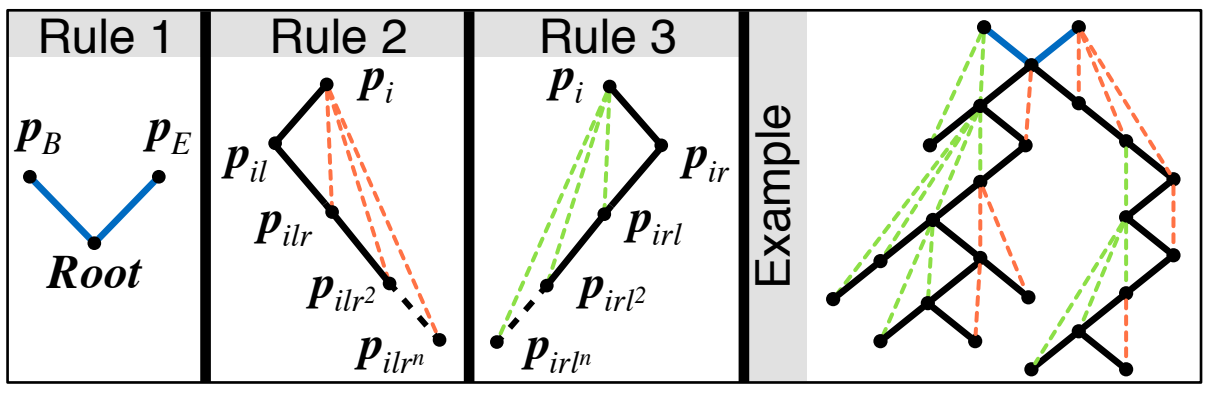

**Figure 4:** *Three rules for extracting the set $\mathcal{T}$ of all possible line segments from a BLG-tree that could be drawn regardless of how the $\varepsilon()$ function is defined. Black (BLG-tree), blue (Rule 1), orange (Rule 2) and green (Rule 3) connections form $\mathcal{T}$ in the example.*

### 3.3. View-dependent Line Simplification

Note, however, that Eq. (2) is only correct if it has a monotonic behavior w.r.t. traversing the BLG-tree top-down, as a node cannot be included in a refined line version without all its ancestors already being included. Furthermore, since we are not in fact explicitly traversing the BLG-trees, but testing individual points, we must enforce this condition in the representation of our attributed line segments.

Based on Eqs. 1 and 2, a point could be included without its ancestors if (i) its error is larger than the one of its ancestors or (ii) $\varepsilon()$ returns a lower threshold than for one of its ancestors. To resolve (i) we conservatively set each node's error to the maximum $\hat{e}$ of its subtree. In Fig. 2, the error $\hat{e}_2$ of node $\boldsymbol{p}_2$ thus becomes 18.9, the largest of its descendants. Case (ii) depends on the camera and is solved by adopting a view-independent error saturation technique [LP01, BPS04, PG07].

Considering $\boldsymbol{p}_2$ and its descendant $\boldsymbol{p}_3$ in Fig. 5(a), the worst case camera position is aligned with the two points, not between them and on the side of the descendant (i.e. $\boldsymbol{p}_3$). In this configuration, it could be that $\hat{e}_2 \not> \varepsilon(d_2)$ but $\hat{e}_3 > \varepsilon(d_3)$ thus causing the ancestor $\boldsymbol{p}_2$ not to be included. To arrive at a view-independent metric, we store for each node the maximal distance $d_{\max}$ to any of its descendants in the line-refinement tree, as illustrated in Fig. 5(b) for $\boldsymbol{p}_2$. Therefore, we replace Eq. (1) by

$$\hat{e}_i > \varepsilon(d_i - d_{\max i}). \tag{3}$$

However, since $d_{\max}$ of a descendant node may still be larger, see also Fig. 5(c), we assign the maximum $\hat{d}_{\max}$ of each subtree to its root. Given the so saturated and maximized errors and distance values, a line segment is visible iff for its generator and splitter points $\boldsymbol{p}_g$ and $\boldsymbol{p}_s$

$$\hat{e}_g > \varepsilon(d_g - \hat{d}_{\max g}) \wedge \hat{e}_s \not> \varepsilon(d_s - \hat{d}_{\max s}). \tag{4}$$

Using Eq. (4) we may conservatively display more details than needed since we consider the worst case configurations. However, it allows us to guarantee a view-dependent LOD approximation error in screen-space.

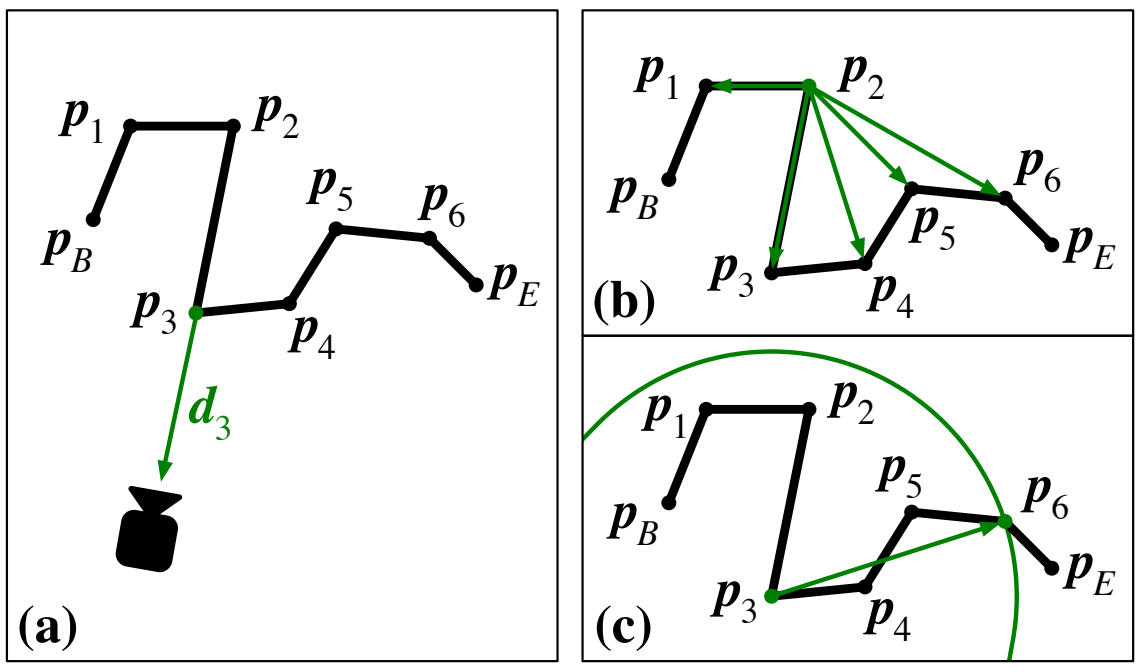

**Figure 5:** *(a) The maximum difference between distances of two points to a camera occurs when the points and the camera are in a line. (b) Calculating the maximum distance* $d_{\max}$ *to descendant nodes for* $\boldsymbol{p}_2$*. (c) Non saturated* $d_{\max}$ *of descendant* $\boldsymbol{p}_3$ *being larger than that of ancestor* $\boldsymbol{p}_2$*.*

### 3.4. Avoiding Line Intersections

Arbitrary line simplifications can cause unwanted intersections of line segments. In Fig. 6, removing $\boldsymbol{p}_1$ and $\boldsymbol{p}_2$ will cause the simplified line $\overline{\boldsymbol{p}_0\boldsymbol{p}_3}$ to intersect another polyline $\overline{\boldsymbol{p}_4\boldsymbol{p}_5\boldsymbol{p}_6}$. This can in fact be predicted by testing wether any other visible point lies inside the triangle formed by the removed point and its baseline, or equivalently the $\triangle_s(\boldsymbol{p}_l, \boldsymbol{p}_g, \boldsymbol{p}_s)$ of line endpoint, generator and splitter points. We call such points, e.g. $\boldsymbol{p}_5$ inside the $\triangle(\boldsymbol{p}_0, \boldsymbol{p}_1, \boldsymbol{p}_3)$ in Fig. 6, *dependees*. Recall the basic exclusion rule of a point $\boldsymbol{p}_i$ being

$$\hat{e}_i \not> \varepsilon(d_i - \hat{d}_{\max i}). \tag{5}$$

Given a refinement or splitter point $\boldsymbol{p}_i$ and its dependees $\mathcal{P}_i = \{\boldsymbol{p}_j | \boldsymbol{p}_j \text{inside} \triangle_i\}$, we must make sure that $\boldsymbol{p}_i$ is excluded only if all points in $\mathcal{P}_i$ are also excluded. Assuming $\mathcal{P}_i$ implicitly includes $\boldsymbol{p}_i$ itself, then we can reformulate Eq. (5) to exclude $\boldsymbol{p}_i$ from refining a line segment to

$$\max_{\boldsymbol{p}_j \in \mathcal{P}_i} \hat{e}_j \not> \min_{\boldsymbol{p}_j \in \mathcal{P}_i} \varepsilon(d_j - \hat{d}_{\max j}). \tag{6}$$

Given the distances $d_i, d_j$ to the camera for the dependent and dependee points $\boldsymbol{p}_i, \boldsymbol{p}_j$ as well as the distance $d_{i,j}$ between them, we know that $d_j \geq d_i - d_{i,j}$. If we plug this into Eq. (6) and reorganize it given that $\varepsilon()$ is a monotonically increasing function, we get

$$\max_{\boldsymbol{p}_j \in \mathcal{P}_i} \hat{e}_j \not> \varepsilon\left(d_i - \max_{\boldsymbol{p}_j \in \mathcal{P}_i} (d_{i,j} + \hat{d}_{\max j})\right). \tag{7}$$

As there is only one term, $d_i$, that varies at run-time in Eq. (7), we can pre-evaluate the remaining terms $e_i^* = \max_{\boldsymbol{p}_j \in \mathcal{P}_i} \hat{e}_j$ and $d_i^* = \max_{\boldsymbol{p}_j \in \mathcal{P}_i} (d_{i,j} + \hat{d}_{\max j})$, and store these two pre-evaluated terms instead of $\hat{e}_i$ and $\hat{d}_{\max i}$ with the point $\boldsymbol{p}_i$.

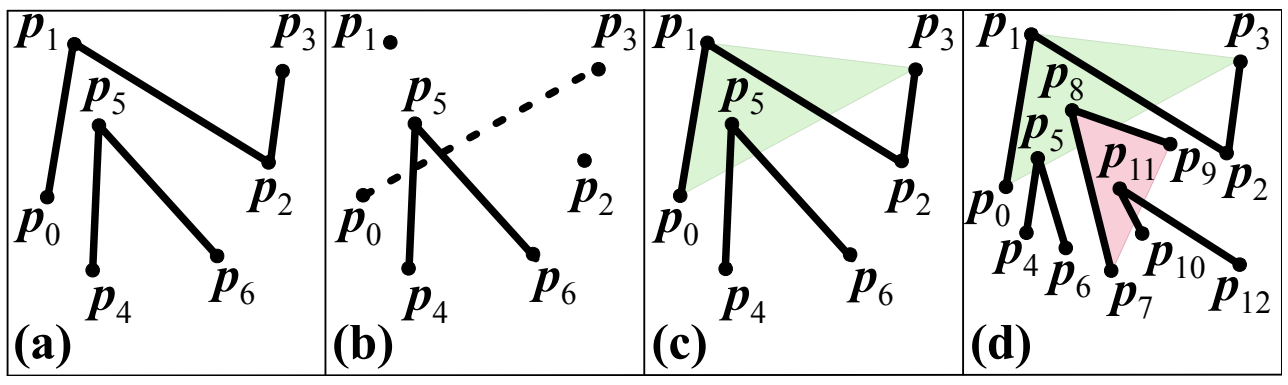

**Figure 6:** *(a, b) An example of two lines that intersect when only one of them is simplified. (c) The intersection happens when* $\boldsymbol{p}_1$ *is excluded and* $\boldsymbol{p}_5$*, its dependee, is not excluded. (d) Exclusion of a point can depend on several dependees and their dependees.*

Eventually, points can have multiple nested dependencies, like in Fig. 6(d), $\boldsymbol{p}_1$ being dependent on $\boldsymbol{p}_5$ and $\boldsymbol{p}_8$ which in turn depends on $\boldsymbol{p}_{11}$. While these linear dependencies can be solved by the saturation mechanism, potential cycles are not. We handle cyclic dependencies by enforcing simultaneous selection of all points in the cycle through the introduction of a proxy point. This proxy point has maximized attributes and an average spatial position and it is used for the evaluation of the exclusion criterion. Though the original point coordinates are still used for drawing. Overlapping cycles are merged and have a single proxy for all their points.

### 3.5. Line Preprocessing Summary

An overview of the preprocess is illustrated in Fig. 7. From the vector map data all points are extracted and stored in a global array. All polylines are transformed into BLG-trees and their nodes' error and distance attributes are saturated as outlined above and are assigned to their respective points. From the BLG-trees a global list of line segments is extracted, each containing four point references: two endpoints, the generator, and the splitter. Now each line segment can independently be evaluated w.r.t. being visible, by evaluating the generator and splitter exclusion, for a given camera position and screen-space error threshold function ε().

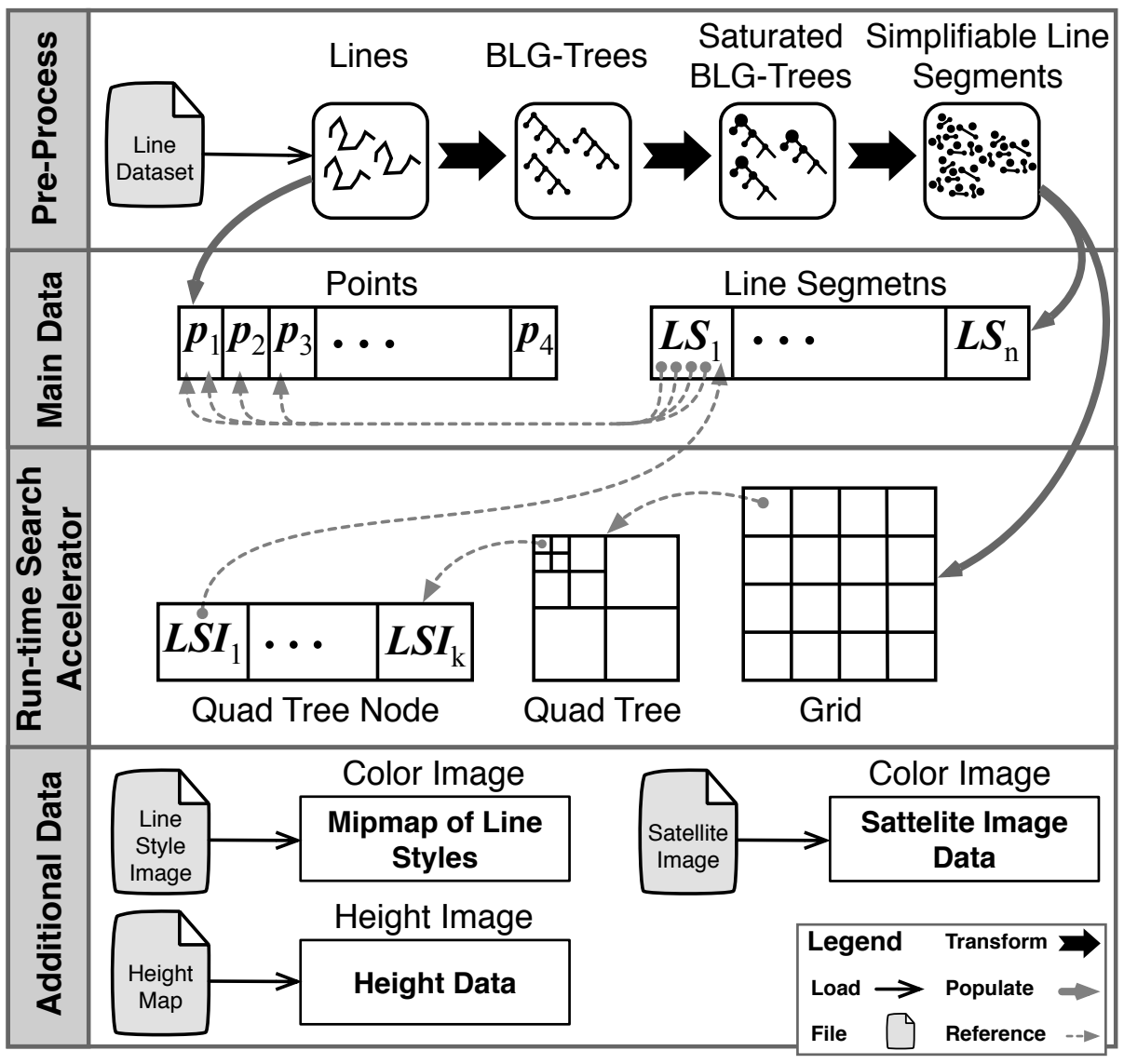

**Figure 7:** *The point attribute and line segment data arrays generated in the preprocess are mapped to a combined grid and quadtree spatial search acceleration data structure on the GPU. Additionally, line styles (see Section 4.2), terrain height and terrain color are loaded for rendering the scene.*

### 3.6. GPU-based Line Visualization

Our interactive line visualization is based on the deferred vector map rendering proposed in [TBP16, TBP18] and extended in [FEP18]. Similar to deferred shading techniques, the actual drawing takes place in a fragment shader that performs pixel-on-line evaluations after the 3D terrain has been rendered in a standard geometry and texturing pass. Through inverse view-projection, each pixel is mapped back into the coordinate system of the vector map data and tested against the relevant and visible line segments. If the pixel is determined to be covered by a line, it is colored and styled accordingly.

This last step, as it is performed for all fragments in the framebuffer in parallel on the GPU, requires a very efficient spatial search index over all line segments. Different spatial search data structures can be used, in [TBP18] a two-level bounding volume hierarchy and in [FEP18] a spatial hash with nested quadtrees is used. We follow a similar two-level principle where a coarse global grid allows quick constant-time access to a cell, see also Fig. 7. In each cell, we use a quadtree for locally refining the search by a branch-and-bound based traversal. We evaluate the visibility of each line segment in the search result using the exclusion criteria of its generator and splitter. If it is visible and its distance to the fragment is less than half the width of the line, as it should appear in screen-space, the corresponding pixel will be colored as indicated by the lines style and pattern.

Where lines meet, consistent joints must be displayed as shown in Fig. 8. Round joints can be achieved by drawing a half-circle at each endpoint using the screen-space distance of the pixel to the line's endpoint. For uniformly colored lines no further special treatment is needed. For styled lines, the minimum distance to both lines' endpoints is used to determine the final color. Pixel $P$ in Fig. 8(g) is located on the black outline of the righthand line and within the grey area of the lefthand line. Since $P$ is closer to the left line, it will be colored in grey. As both lines are needed for coloring pixels near joints, we continue the search after hitting the first line to find the second line.

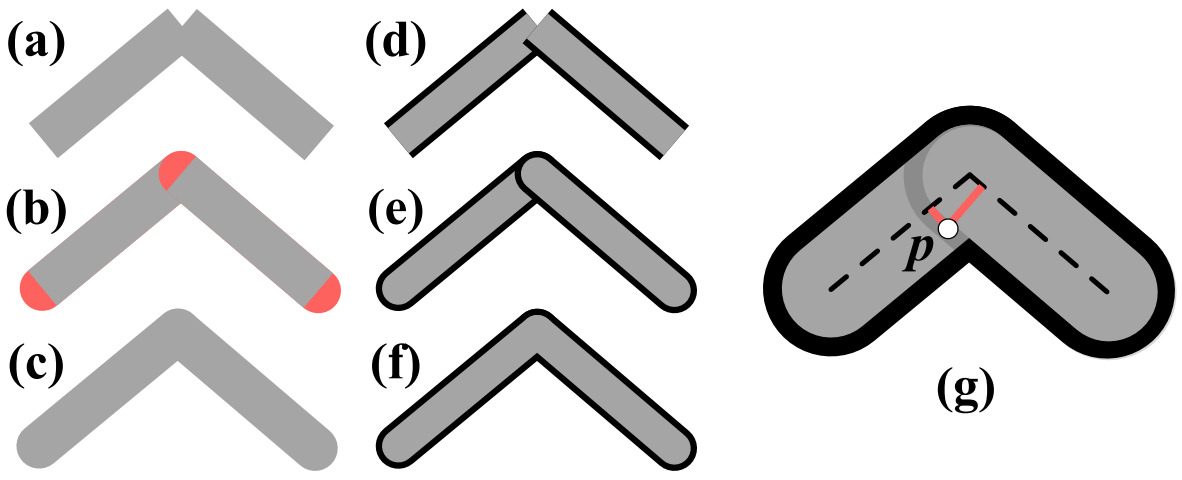

**Figure 8:** *(a) Separate line segments disconnected at a joint. (b) A simple solution draws a half-circle at each end. (c) Uniformly colored line segments appear continuous with overlapping half-circles. (d, e) For styled lines, half-circles do not solve the problem completely. In order for joints to look correct (f), we must use the color of the line that is closer to the pixel (g).*

## 4. Implementation

### 4.1. Deferred Line Rendering

We have implemented our approach in C++ using OpenGL4.1 to limit dependency on advanced graphics features and thus support a wider range of possible applications. In the preprocess, a line dataset in *shapefile* format is transformed into three data textures for the rendering application, containing the points, line segments and the spatial search grid and line segment quadtrees, see also Fig. 7. The renderer uses these three data textures along with a terrain heigh-field mesh, an image texture for the terrain, and a style-texture for the line categories. During rendering, the terrain elevation model with its texture(s) is processed by a regular textured geometry rendering pass, exploiting multiresolution terrain rendering techniques. The vector map line rendering actually takes place solely in the fragment shader, which not only performs the normal terrain shading and texturing after vertex transformation, but also executes the per-pixel line evaluations.

From the world coordinates provided by the vertex shader, similar to [TBP18], OpenGL generates the exact world coordinates for all fragments. Using per-pixel world coordinates, the fragment shader finds the grid cell in which the fragment is located and traverses the corresponding line segment quadtree, calculating the distance between the fragment and each line segment that it encounters. Given the distance, line width and pixel size, a percentage area coverage is computed, used for antialiased blending as detailed below. For any non-zero overlap, the pixel's color is set to the line type's style texture based on its distance to the line segment. Further overlapping line segments go through the same process if they have the same or higher line type priority, while lower priorities will be dropped. In case of higher priority, the fragment color will be replaced, in case of equal priority the color of the closer line segment is kept. The latter case also includes handling joints as described in Section 3.6. Alg. 1 describes the process in more details.

The number of texture lookups strongly affects the performance of the fragment shader. Since each access can return a RGBA color value or XYZW homogeneous coordinate 4-tuple, we structured our textures in a way to reduce the number of lookups. The first two values of each tuple in the points texture hold the $x, y$ world coordinates of the point, the other two are the saturated error $e^*$ and distance $d^*$ values. For each line segment, we need the indices of three different points, i.e. two endpoints and the splitter, which are stored in the first three values. Additionally, the first bit of the last value indicates which endpoint is the line's generator, and the remaining bits are used for storing the type of the line. The line segment quadtree texture contains two types of tuples: address and data tuples. An address tuple holds the indices of the four child nodes. Multiple consecutive data tuples store the indices of the line segments inside a node.

### 4.2. Antialiasing

We use two techniques to overcome staircase and minification aliasing artifacts that appear when perspectively projecting and discretizing lines onto a fixed resolution screen. Note that magnification artifacts do not occur in our deferred line rendering method as we are basically doing a pixel-precise rasterization in the fragment shader which never causes a magnification.

Staircase aliasing at the outer edges of lines is prevented by using the per-pixel area coverage of the pixel-line overlap as the alpha value of the selected line color. In order to avoid aliasing inside the lines that have style patterns, we exploit the OpenGL mipmapping functionality. Fig. 9 shows a line style texture that stores nine types of styles at several different mipmap levels. At the top mipmap level (0), each style type covers a $512 \times 256$ pixel matrix which results in a $512 \times 2304$ texture in total. Since the style block is $256 = 2^8$ wide, a mipmap pyramid of nine levels can be created without mixing the colors of adjacent styles.

Furthermore, given the alpha-blending based antialiasing we can further exploit this for adjusting the line thickness dynamically. Our approach allows to increase or decrease the line thickness adaptively based on distance, as well as other spatial or even temporal functions. In our current implementation, we progressively increase the thickness of important lines and reduce it for others by distance. Therefore, small lines smoothly become invisible at far distances due to becoming subpixel in size while other important ones remain visible. As the camera gets closer, this distance based adjustment is cancelled out and all lines are gradually adjusted to their actual thickness. We need to assign the line segments to the quadtree nodes that they cover at their largest thickness. We multiplied the base thickness of each line type by a factor that we determined practically. This effect is applied in Fig. 10, resulting in a much less cluttered view of far areas as in the top image.

**Algorithm 1** Fragment Shader

```
Input fragmentWorldCoordinate(fWCoord), pixelSize,
      lineStyle, points, lines, quadTrees, terrainColor
Output fragmentColor
function CALCULATEFRAGMENTCOLOR
    currentNode ← root of quadTree containing fWCoord
    coverage ← 0
    currentPriority ←lowest priority
    currentDistance ← ∞
    while currentNode is not empty do
        for all lines in currentNode do
            if line is generated and is not split then
                coverage ← CoveredByLine(fWCoord,
                 line, pixelSize)
                distance ← distance(fWCoord, line)
                if coverage > 0 and
                 priority of line > currentPriority or
                 (priority of line = currentPriority and
                 distance < currentDistance) then
                    fragmentColor ← readLineStyleColor(
                     type of line, distance)
                    alpha of fragmentColor ← coverage
                    currentPriority ←priority of line
                    currentDistance ← distance
                end if
            end if
        end for
        currentNode ← child node of currentNode that
         contains fWCoord
    end while
    fragmentColor ← blend(terrainColor, fragmentColor)
    return fragmentColor
end function
```

## 5. Results

We tested our system on two different computers: a 4GHz Intel Core i7-6700K, 16GB RAM, AMD Radeon R9 M395x running MacOS (SYS1) and a 3.5GHz Core i7-3770K, 16GB RAM, GeForce GTX 1080Ti running Windows (SYS2). We used three different data layers for our experiments: (1) a heightmap for creating the terrain mesh, (2) a terrain texture provided by Swisstopo, and (3) a street dataset. We used two street datasets: Open Street Map (DS1) and Swisstopo VECTOR25 (DS2). Tab. 1 contains more information about these datasets. Due to being hardly distinguishable visually, only screenshots of DS1 are depicted in this section. The resolution of the frame buffer was $1920 \times 1080$ on both systems in all experiments. In this section, we discuss our results from three view points: rendering, simplification, and performance.

### 5.1. Pixel-Precise Line Rendering

Our software can successfully load a large-scale vector map dataset and project it on a large terrain. It can deliver pixel-precise line rendering without any pixelation artifacts irrespective of the zoom factor and without any recognizable aliasing artifacts. Although we used a simple technique with no LOD management for rendering the 3D terrain, our implementation is independent of the terrain rendering itself and proves that it is capable of blending the lines with the textured terrain seamlessly. Fig. 10 presents multiple screenshots at various zoom levels, displaying the whole street network of Switzerland at highest LOD. As we zoom in, the lines become less cluttered and their style becomes clearly recognizable as soon as the line width covers several pixels.

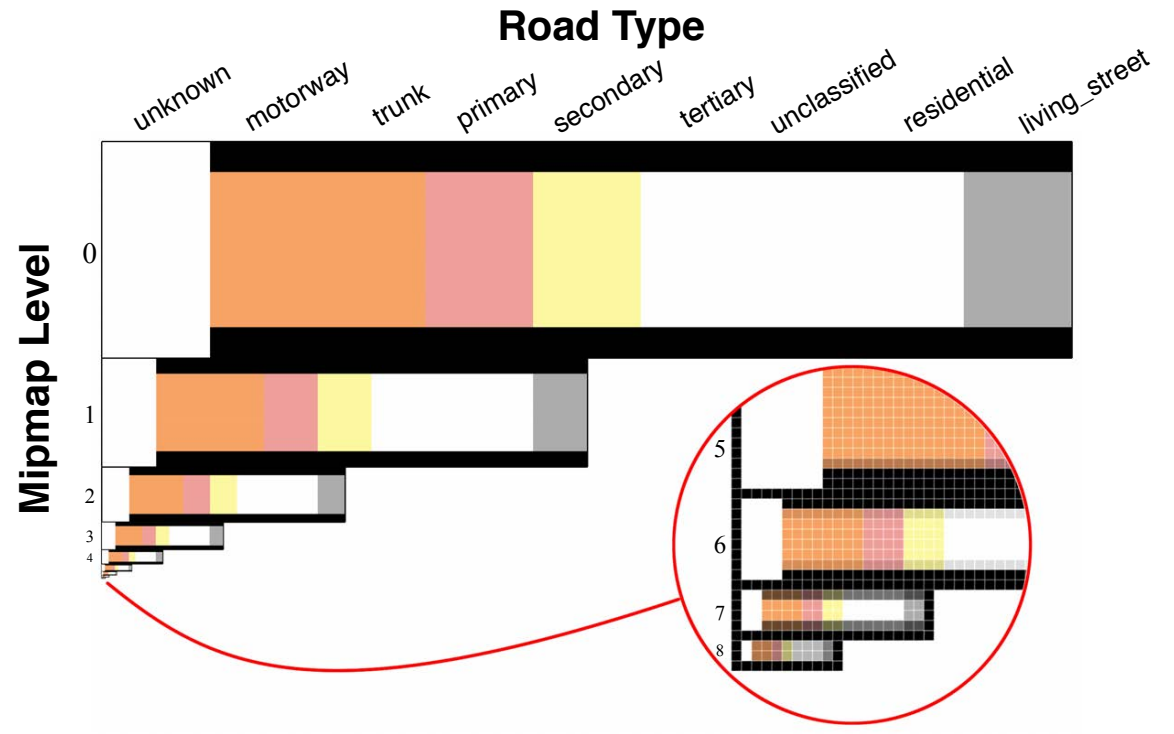

**Figure 9:** *Mipmap pyramid of line style texture. The texture contains nine line styles. In the magnified section, the pixels can be distinguished by their white borders. In level eight, each row of the texture containing two pixels represent a line style.*

| | # points | Mem. for points | # segs | # ap-segs | Mem. for ap-segs | # qt-segs | Mem. for qt-segs | # dt-qt-segs | Mem. for dt-qt-segs |
|---|---|---|---|---|---|---|---|---|---|
| **DS1** | 7.7M | 235.7MB | 6.9M | 13M | 198.6MB | 47.4M | 1.3GB | 83.3M | 1.8GB |
| **DS2** | 12.4M | 380.8MB | 11.1M | 21M | 320.2MB | 75.1M | 2.1GB | 125.1M | 2.9GB |

**Table 1:** *Information of datasets DS1 (Open Street Map) and DS2 (Swisstopo VECTOR25) including the number of points, segments (segs), all-possible segments (ap-segs), segments assigned to the quadtrees (qt-segs), segments assigned to the quadtrees when dynamic thickness is enabled (dt-qt-segs), and the memory they need.*

### 5.2. Adaptive Line Simplification

Our locally-adaptive line simplification technique, LOCALIS, demonstrates that real-time line simplification can be applied to interactive vector map visualization applications. The approach includes a pre-processing effort, which is performed once offline, and a line data storage cost for managing all line segments that can possibly appear when refining the lines. The impact of this cost is twofold. First, it requires some extra memory to store the line segments which, however, has not been a bottleneck in our current implementation. Second, handling large-scale vector map line data has only recently been made possible, and doubling the number of line segments may thus have a limiting effect on the performance of such techniques. Performance is discussed in the following section.

The real-time adaptive line simplification of LOCALIS is shown in Fig. 11 by highlighting areas of line-refinement. Using a lens tool, outlined by the red circle, the polylines are interactively refined inside and simplified outside. A recorded interactive demonstration is supplied in the accompanying supplemental video.

### 5.3. Performance

We identified four parts of LOCALIS that chiefly affect the performance: the dynamic line thickness, the visibility evaluation and the increase in the number of lines as a result of creating all possible lines. We designed four experiments to demonstrate the influence of each part: (1) AVD: All possible lines, Visibility check, and Dynamic line thickness. (2) AVS: All possible lines, Visibility check, and Static line thickness. (3) ANVS: All possible lines, No Visibility check, and Static line thickness. (4) ONVS: Original line segments, No Visibility check, and Static line thickness. AVD has all features of LOCALIS while ONVS is equivalent to just rendering the lines. This supports a performance comparison of the LOCALIS specific features to a base-line configuration.

We ran our performance tests on datasets DS1 and DS2. A detailed information about these datasets is given in Tab. 1. We used the datasets without preprocessing since they did not have connections at middle points of the lines. In the left column of Fig. 12, four snapshots at different zoom levels are depicted. In the right column, the corresponding heatmaps show the number of distance-to-line tests per fragment. In Fig. 13, the amount of time needed to render the terrain and the street lines are shown in milliseconds as bar charts. For each zoom level, four experiments are done with two datasets on two machines creating 16 bars.

In our case, run-time performance depends on three factors: (1) number of lines inspected per fragment and in total each frame, (2) memory locality of the data on the GPU, and (3) LOD threshold. The performance is higher when the majority of the fragments are covered by less populated nodes of the quadtrees, a limited part of the memory is accessed (e.g., in close-up views), and lower LODs are used. The first factor is the most influential. In Fig. 13, we can see the effect of the second factor at Zoom 8, where most of the fragments are covered by dense nodes but the performance is higher than in Zoom 2. The effect of the third factor is not as significant as the other two since all relevant line segments are queried regardless of the LOD threshold and only the number of immediately discarded line segments increases in lower LODs (see Section 3).

Based on the results of ONVS, our base vector line renderer performs at the level of a state-of-the-art technique [TBP16, TBP18] and scales as expected when comparing to ANVS, which deals with twice the number of line segments. The effect of the visibility check is negligible according to AVS. The results of AVD show that the effect of dynamic line thickness on performance is significant. This is due to the widened line segments overlapping extensively at curves, resulting in dense quadtree nodes. This negative impact is subtle in Zoom 6 where no highways or major roads are visible.

To overcome aliasing artifacts we employed two techniques as outlined in Section 4. Fig. 14 shows a view with different types of roads. The magnified insets demonstrate that both techniques successfully smooth the outer edge as well as the inner part of the styled lines and prevent any aliasing artifacts. The smoothed staircase on the outer edge is achieved by precise calculation of the pix-els' area-coverage, and the smooth interior is obtained by querying the line style texture at an appropriate mipmap level. The antialiasing is consistent in lines with different angles and types.

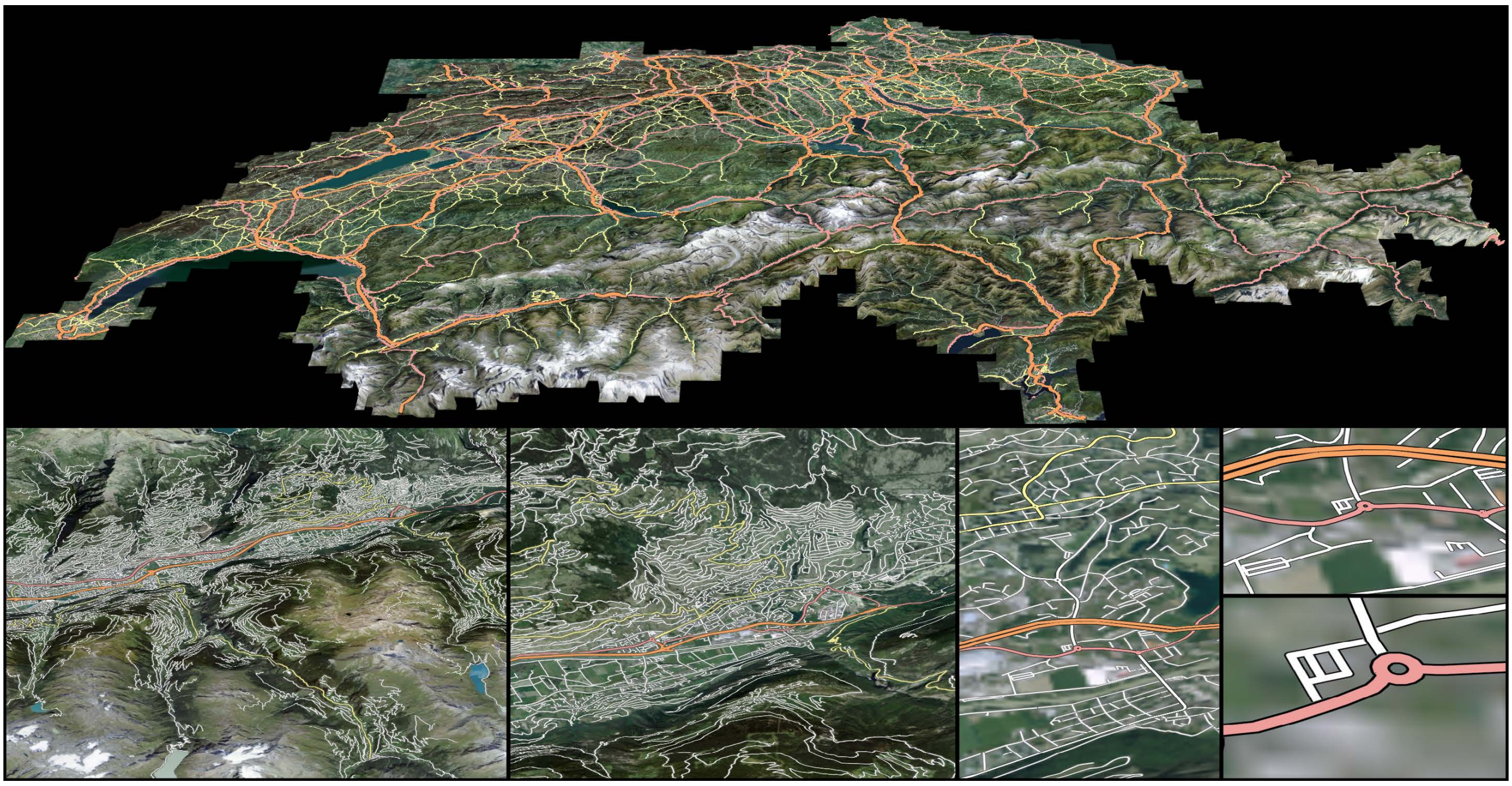

**Figure 10:** *Screenshots of our system visualizing the whole street network of Switzerland over a terrain mesh at full detail. The LOD is set to the highest for demonstration purpose.*

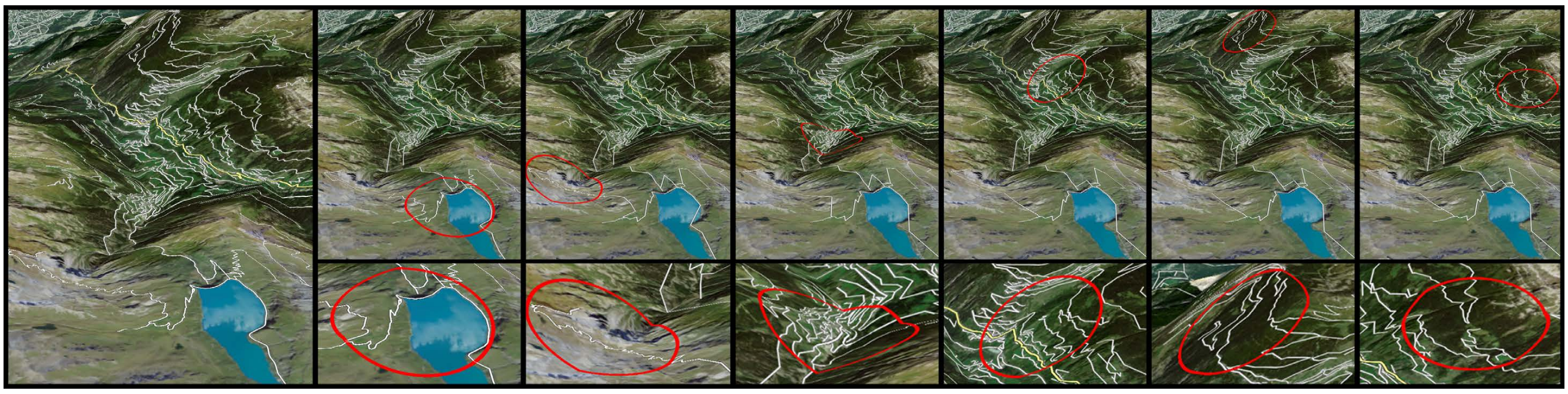

**Figure 11:** Left*: Full detail lines without any simplification are drawn.* Top row*: Sequential snapshots of a line-refinement lens moving interactively over the terrain, with all lines outside the lens being heavily simplified.* Bottom row*: Close-up of the lens in each snapshot, magnifying the refined and simplified lines in- and outside of the lens respectively.*

## 6. Conclusions

In this paper we have presented LOCALIS, our new *locally-adaptive line simplification* technique for simple polylines based on the DP algorithm. Our technique creates every possible line segment that can emerge during line refinement using BLG-trees and makes them individually processable by attributing them. LOCALIS exploits the direct access to line data on the GPU as used by deferred vector map rendering and decides whether a line segment should be displayed based on a given LOD threshold in the last step of the rendering pipeline. Our implementation shows that LOCALIS can always produce and display a pixel-precise and valid simplified representation of the lines regardless of the distribution of the required LOD over the screen. It can simplify any part of a line while keeping the details of the other part.

We integrated LOCALIS with a state-of-the-art deferred vector map rendering algorithm using data structures that serve both algorithms. We have tested our prototype on the whole street network of Switzerland rendered on top of a 3D terrain mesh. In this prototype, the user can manipulate the LOD by moving the camera closer to or farther from the terrain in an arbitrary perspective or by activating a moving virtual line-refinement lens over the terrain. Our implementation shows that line simplification can be done interactively at the cost of an increased number of line segments to be maintained in our data structure.

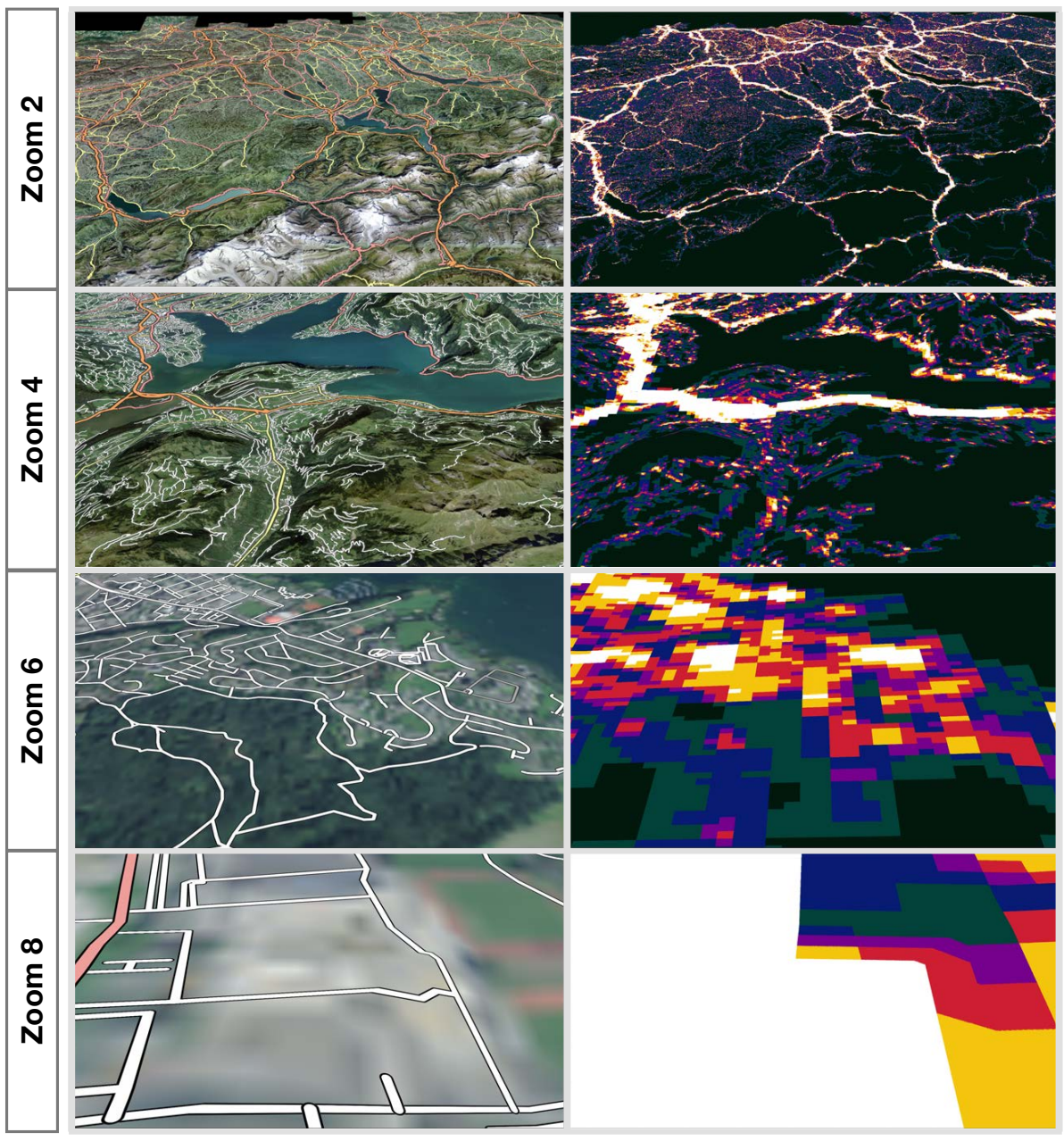

**Figure 12:** *Snapshots at different zoom levels: Left shows the rendered lines at the highest LOD, the right shows the heatmaps indicating the number of distance-to-line tests per pixel.*

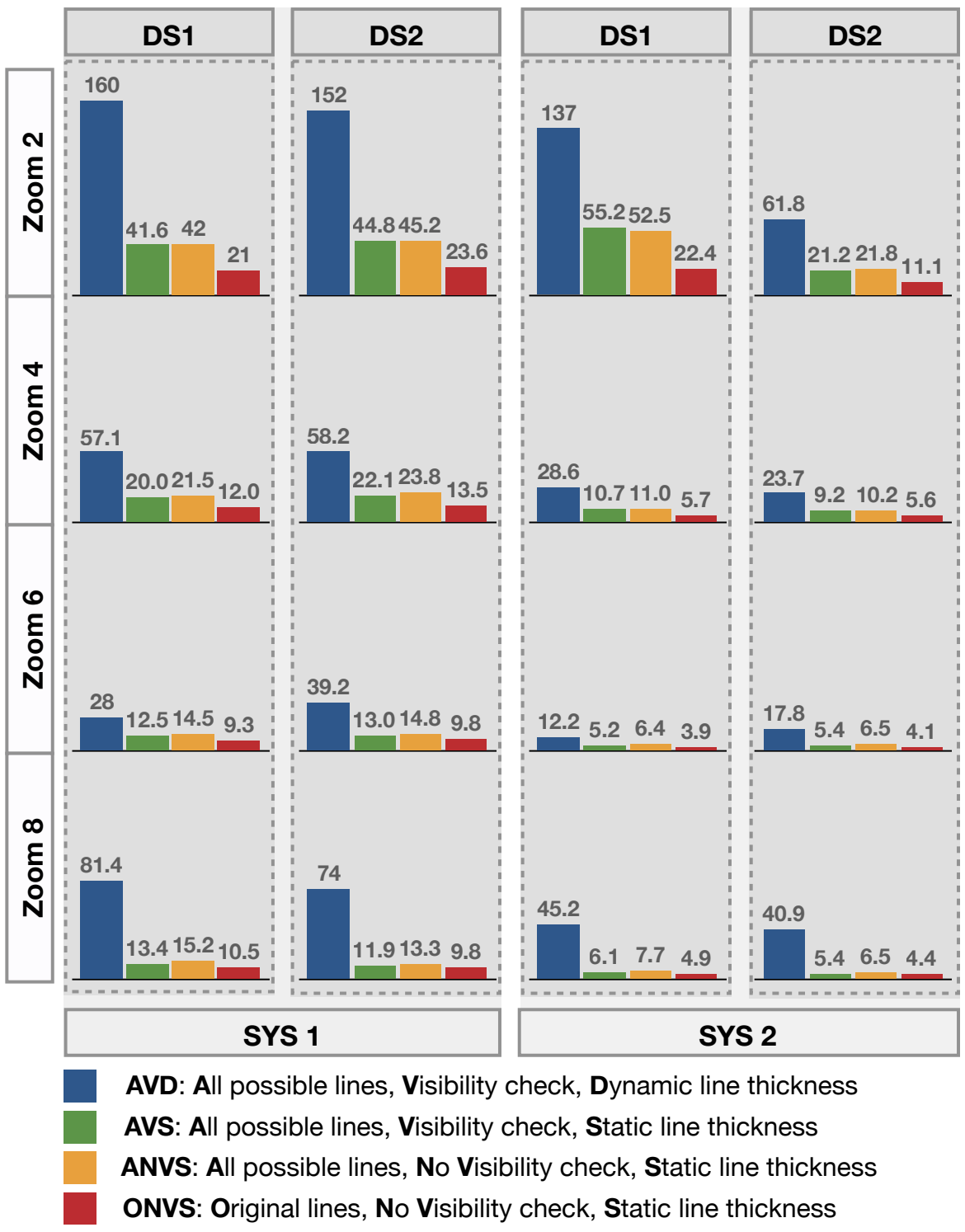

**Figure 13:** *Results of four Experiments with LOCALIS (AVD, AVS, ANVS, and ONVS) rendering two datasets (DS1 and DS2) on two machines (SYS1 and SYS2) at four zoom levels in milliseconds.*

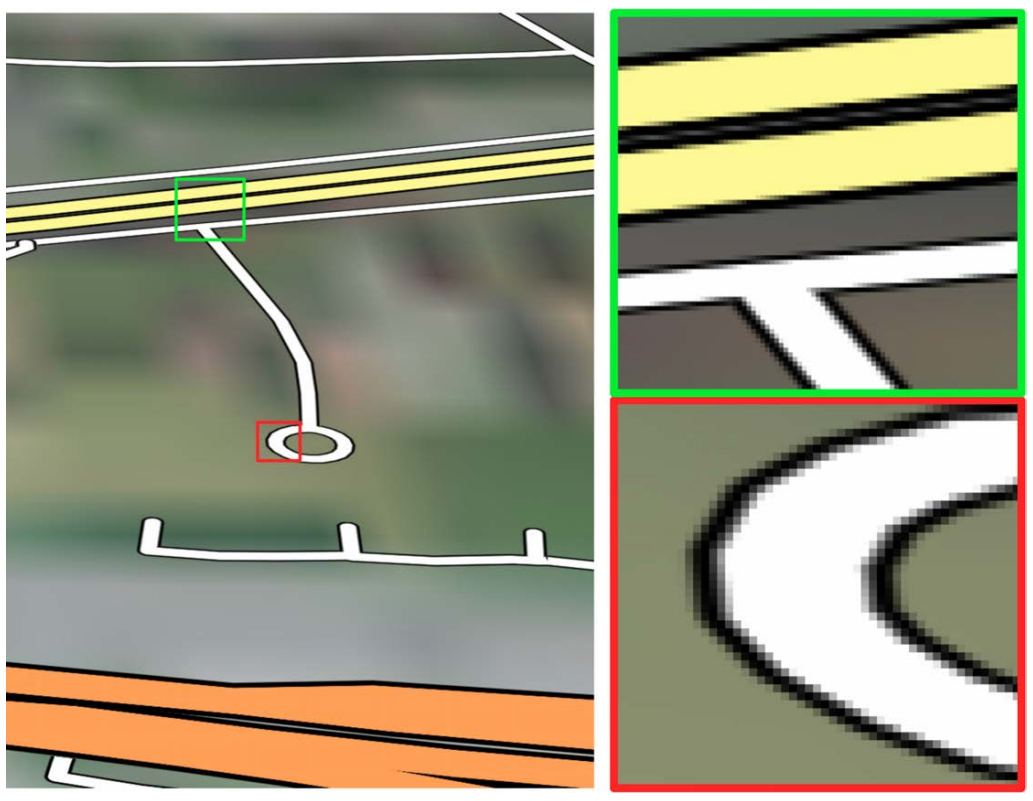

**Figure 14:** *Two regions of the output image (left) are magnified (right). The smooth borders are visible around, and the smoothed out interior within the styled lines.*

In our line simplification approach, we took the characteristics of the target 3D visualization system into account. Knowing that the lines will be displayed in an interactive 3D environment, implies that the lines must be processed in real-time such that the camera can move freely. Our Technique enables applications to dynamically manage the LOD of polylines interactively based on the situation and user demands without needing other proxy data structures such as auxiliary geometry, image textures, or pre-calculated datasets with discrete LODs.

Regarding the scalability, the bottleneck is the amount of available memory. For larger datasets, it would not be possible to load the whole data structures, and further measures are required for loading the parts that are needed based on the viewing angle and LOD. Additionally, the hierarchical data structures can potentially limit the run-time search for lines that cover a pixel, if the depth in which a line is stored is chosen based on the LODs that contain that line. These two points and designing locally-adaptive LOD management algorithms for other types of vector data such as closed polygons and meshes can be done in future work.

## Acknowledgements

The authors want to thank the Swiss Federal Office of Topography Swisstopo for providing the Swiss VECTOR25 and SwissTLM data sets as well as the OpenStreetMap Foundation for access to their data. This project was partially supported by a Swiss National Science Foundation (SNSF) research grant (project no. 200021_169628).